\documentclass{article}      

\usepackage[utf8]{inputenc}
\usepackage[english]{babel}

\usepackage{gensymb}
\usepackage{amssymb}
\usepackage[intlimits]{amsmath}

\usepackage{graphicx}
\usepackage{amssymb}
\usepackage{amsmath}
\usepackage{amsthm}

\usepackage[english]{babel}
\usepackage{epstopdf}
\usepackage{booktabs}

\usepackage[authoryear, sort, round]{natbib}

\usepackage{caption}
\usepackage{subcaption}

\usepackage{natbib}

\usepackage{natbib}

\title{Finite population inference for skewness measures}  
\author{Leo Pasquazzi\footnote{email: leo.pasquazzi@unimib.it, ORCID https://orcid.org/0000-0002-2467-2667}\\
Dipartimento di Statistica e Metodi Quantitativi,\\
Università degli Studi di Milano - Bicocca, \\
Piazza dell'Ateneo Nuovo, 1 - 20126, Milano, Italy}      

\begin{document}             

\maketitle                   

\section*{Abstract}
In this article we consider Bowley's skewness measure and the Groeneveld-Meeden $b_{3}$ index in the context of finite population sampling. We employ the functional delta method to obtain asymptotic variance formulae for plug-in estimators and propose corresponding variance estimators. We then consider plug-in estimators based on the H\'{a}jek cdf-estimator and on a Deville-Särndal type calibration estimator and test the performance of normal confidence intervals.

\bigskip

\noindent \textbf{Keywords:} skewness, Bowley's index, Groeneveld-Meeden $b_{3}$ index, finite population inference

\bigskip


\section{Introduction}

In order to allow for asymmetry and other deviations from normality, \citet{Pearson_1895} introduced his well-known distribution family and proposed 
\[\zeta:=\frac{\text{mode-mean}}{\text{standard deviation}}\]
as a measure for asymmetry. Curiously, only for Type III distributions (location scale transformations of gamma distributions) Pearson uses the difference between the mean and the mode in the numerator rather than the other way round. As Pearson considered the method of moments to fit his distributions ("generalized probability curves") to empirical data, he provided formulae to compute $\zeta$ from the standardized third central moment, i.e. from $\mu_{3}/\sigma^{3}$. Despite its limitations (e.g. nonexistence of moments, large sample variability), $\mu_{3}/\sigma^{3}$ became very popular in the aftermath and is still widely used today.

Some years after the publication of Pearson's paper, \cite{Bowley_1901} proposed 
\[\frac{\text{upper interquartile range - lower interquartile range}}{\text{central interquartile range}}\]
as a measure for skewness. Actually, in the first edition of his 1901 book "Elements of Statistics", Bowley did not write down the explicit definition of this measure, but the underlying idea can be inferred from some comments on an example about the distribution of wages among branches of the Amalgameted Society of Engineers \citep[see pages 134--136 in][]{Bowley_1901}. For sure the explicit definition of Bowley's measure appears in the fourth edition of his book which was published in 1920 \citep[see page 116 in][]{Bowley_1920}. Perhaps it appears also in the second and/or third edition which where published in 1902 and 1907, respectively, but I was not able to verify this since I had no access to these intermediate editions. 

For estimation purposes in the Box-Cox model, \citet{Hinkley_1975} considers an immediate generalization of Bowley's measure which is given by
\[b_{2}(r):=\frac{F^{-1}(1-r)+F^{-1}(r)-2F^{-1}(0.5)}{F^{-1}(1-r)-F^{-1}(r)},\quad r\in(0,1),\]
where $F^{-1}(r):=\inf\{x\in\mathbb{R}: F(x)\geq r\}$ denotes the $r$-quantile of the distribution function $F(x)$. Of course, $b_{2}(r)$ coincides with Bowley's measure when $r=0.25$. Finally, in close relation to $b_{2}(r)$, \citet{Groeneveld_Meeden_1984} propose
\[b_{3}:=\frac{\int_{0}^{1}\{F^{-1}(1-r)+F^{-1}(r)-2F^{-1}(0.5)\}dr}{\int_{0}^{1}\{F^{-1}(1-r)-F^{-1}(r)\}dr}=\frac{\mu-\nu}{E(|X-\nu|)}\]
as a global measure for skewness. Here, we used $X$ to denote a random variable with distribution function $F(x)$, $\mu$ to indicate $E(X)$ and $\nu$ to indicate the median $F^{-1}(0.5)$. As shown in \citet{Groeneveld_Meeden_1984}, $b_{2}(r)$ and $b_{3}$ are both consistent with the partial skewness ordering relation of \cite{vanZwet_1964}.

The present article is about estimation of $b_{2}(r)$ and $b_{3}$ in the context of finite population sampling. In section 2 we introduce plug-in estimators based on the H\'{a}jek cdf-estimator and on a Deville-Särndal type calibration estimator. We also provide asymptotic variance formulae and corresponding variance estimators. In section 3 we test the estimators in a simulation study. Normal confidence intervals are also tested. Conclusions and final remarks end this paper in section 4.

\section{Estimators, asymptotic variance and variance estimators}

Let $U_{N}:=\{1,2,\dots,N\}$ be the set of labels which identify the units of a finite population, let $y_{1}$, \dots, $y_{N}$ be the values taken on by a study variable $Y$ and let $x_{1}$, \dots, $x_{N}$ be the values taken on by an auxiliary variable $X$. The finite population cdf of $Y$ is given by $F(t):=\frac{1}{N}\sum_{i=1}^{N}I(y_{i}\leq t)$ and henceforth we assume that $b_{2}(r)$ and $b_{3}$ refer to $F(t)$. 

Now, let $s:=\{i_{1}, i_{2}, \dots, i_{n}\}\subset U_{N}$ be a sample taken from $U_{N}$ ($n$ denotes the sample size). In this article we assume the $X$-values to be known for every unit $i\in U_{N}$, but that the $Y$-values be known only for $i\in s$. For $i,j\in U_{N}$ we denote the first and second order sample inclusion probabilities by $\pi_{i}$ and $\pi_{ij}$, respectively.

As already mentioned above, in this article we consider plug-in estimators based on
\begin{itemize}
\item[a) ] the H\'{a}jek estimator 
\[\widehat{F}_{Ha}(t):=\frac{1}{\sum_{i\in s}\frac{1}{\pi_{i}}}\sum_{i\in s}\frac{1}{\pi_{i}}I(y_{i}\leq t):=\frac{1}{\widehat{N}}\sum_{i\in s}\frac{1}{\pi_{i}}I(y_{i}\leq t)\]
\item[b) ] and the calibration estimator \cite[see][]{Deville_Sarndal_1992}
\[\widehat{F}_{cal}(t):=\frac{1}{N}\sum_{i\in s}w_{i}I(y_{i}\leq t)\]
with sample weights $w_{i}$ given by
\begin{equation}\label{further_weights}
w_{i}:=\frac{\exp(\widehat{\beta}_{0}+x_{i}\widehat{\beta}_{1})}{\pi_{i}}
\end{equation}
where $\widehat{\beta}:=[\widehat{\beta}_{0}, \widehat{\beta}_{1}]$ solves the calibration equations
\begin{equation}\label{condizione_beta}
\begin{split}
\sum_{i\in s}\frac{\exp(\widehat{\beta}_{0}+x_{i}\widehat{\beta}_{1})}{\pi_{i}}&=N\\
\sum_{i\in s}\frac{\exp(\widehat{\beta}_{0}+x_{i}\widehat{\beta}_{1})}{\pi_{i}}x_{i}&=\sum_{i=1}^{N}x_{i}.\\
\end{split}
\end{equation}
which are always solvable.
\end{itemize}
Note that for some widely used sample designs (e.g. simple random sampling and stratified simple random sampling) $\widehat{F}_{Ha}(t)$ reduces to the Horvitz-Thompson estimator 
\[\widehat{F}_{HT}(t):=\frac{1}{N}\sum_{i\in s}\frac{1}{\pi_{i}}I(y_{i}\leq t).\]
Also note that $\widehat{F}_{Ha}(t)$ and $\widehat{F}_{HT}(t)$ do not account for the auxiliary information unless the sample design does so through the first order inclusion probabilities $\pi_{i}$. When the latter is not the case, it seems more appealing to use $\widehat{F}_{cal}(t)$ instead of $\widehat{F}_{Ha}(t)$ or $\widehat{F}_{HT}(t)$. 

Now, consider the plug-in estimators. The ones corresponding to $\widehat{F}_{Ha}(t)$ will be denoted by $\widehat{b}_{2,Ha}(r)$ and $\widehat{b}_{3,Ha}$, respectively, while those corresponding to $\widehat{F}_{cal}(t)$ will be denoted by $\widehat{b}_{2,cal}(r)$ and $\widehat{b}_{3,cal}$. Appendix A provides an argument by which
\[\frac{\widehat{b}_{\bullet, \diamond}-b_{\bullet}}{V_{N,\bullet,\diamond}}\overset{\mathcal{L}}{\longrightarrow}N(0,1)\]
under broad conditions, where $\bullet$ is either "$2$" or "$3$", $\diamond$ is either "$Ha$" or "$cal$" and $V_{N,\bullet,\diamond}^{2}$ is the sequence of asymptotic variances to be defined below. Of course, when $\bullet=2$, $\widehat{b}_{\bullet, \diamond}$ and $V_{N,\bullet,\diamond}^{2}$ depend also on $r$ but this dependence does not show up in the notation. From formulae (\ref{asymptotic_variance_sequence}) and (\ref{asymptotic_variance_sequence_1}) in appendix A we know that 
\[V_{N,\bullet,\diamond}^{2}=\frac{1}{N^{2}}var\left(\sum_{i\in s}d_{i}g(y_{i})\right)\]
where
\begin{equation*}
d_{i}=\begin{cases}
N/(\widehat{N}\pi_{i}) & \text{ if }\diamond="Ha",\\
w_{i} \text{ defined in (\ref{further_weights}) and (\ref{condizione_beta})}& \text{ if }\diamond="cal".\\
\end{cases}
\end{equation*}
and where the function $g(t)$ is defined by either (\ref{g_3_definition}) or (\ref{g_2_definition}) in appendix A according to whether $\bullet$ is "3" or "2". 

\smallskip
Now, consider the case where $\diamond="Ha"$. A straightforward application of the Delta-method yields
\[\sum_{i\in s}d_{i}g(y_{i})=\sum_{i\in s}\frac{N}{\widehat{N}\pi_{i}}g(y_{i})\simeq N\overline{g}+\sum_{i\in s}\frac{1}{\pi_{i}}[g(y_{i})-\overline{g}]\]
where $\overline{g}:=\frac{1}{N}\sum_{i=1}^{N}g(y_{i})$. Hence we obtain
\[V_{N,\bullet,Ha}^{2}\simeq var\left(\sum_{i\in s}\frac{1}{\pi_{i}}[g(y_{i})-\overline{g}]\right)=\sum_{i=1}^{N}\sum_{j=1}^{N}\frac{\pi_{ij}-\pi_{i}\pi_{j}}{\pi_{i}\pi_{j}}[g(y_{i})-\overline{g}][g(y_{j})-\overline{g}].\]
As estimator for $V_{N,\bullet,Ha}^{2}$ we may consider the standard Horvitz-Thompson variance estimator
\[\widehat{V}_{N,\bullet,Ha,HT}^{2}:=\sum_{i\in s}\sum_{j\in s}\frac{\pi_{ij}-\pi_{i}\pi_{j}}{\pi_{ij}\pi_{i}\pi_{j}}[\widehat{g}(y_{i})-\overline{\widehat{g}}][\widehat{g}(y_{j})-\overline{\widehat{g}}],\]
where $\widehat{g}(t)$ is an estimate for the unknown function $g(t)$ (below we propose such an estimate). However, it is well known that for fixed size designs the Sen-Yates-Grundy estimator
\[\widehat{V}_{N,\bullet,Ha,SYG}^{2}:=-\frac{1}{2}\sum_{i\in s}\sum_{j\in s}\frac{\pi_{ij}-\pi_{i}\pi_{j}}{\pi_{ij}}\left(\frac{\widehat{g}(y_{i})-\overline{\widehat{g}}}{\pi_{i}}-\frac{\widehat{g}(y_{j})-\overline{\widehat{g}}}{\pi_{j}}\right)^{2}\]
is usually preferable \citep[see][]{Vijayan_1975}. In the simulation study of this article, where only fixed size designs are considered, we used the latter.

\smallskip
Next consider the case where $\diamond="cal"$. In this case we have \citep[see][]{Deville_Sarndal_1992}
\[V_{N,\bullet,cal}^{2}\simeq \frac{1}{N^{2}}var\left(\sum_{i\in s}w_{i}g(y_{i})\right)=\sum_{i=1}^{N}\sum_{j=1}^{N}(\pi_{ij}-\pi_{i}\pi_{j})w_{i}w_{j}E_{i}E_{j}\]
where the $E_{i}$'s are the residuals of the least squares regression of the $g(y_{i})$'s on the $x_{i}$'s and a constant. A corresponding Horvitz-Thompson-type variance estimator is given by
\[\widehat{V}_{N,\bullet,cal,HT}^{2}:=\sum_{i\in s}\sum_{j\in s}\frac{\pi_{ij}-\pi_{i}\pi_{j}}{\pi_{ij}}w_{i}w_{j}\widehat{E}_{i}\widehat{E}_{j},\]
where the $\widehat{E}_{i}$'s are the residuals of the $w_{i}$-weighted least squares regression of the sample $\widehat{g}(y_{i})$'s on the sample $x_{i}$'s and a constant. The corresponding Sen-Yates-Grundy estimator is given by
\[\widehat{V}_{N,\bullet,cal,SYG}^{2}:=-\frac{1}{2}\sum_{i\in s}\sum_{j\in s}\frac{\pi_{ij}-\pi_{i}\pi_{j}}{\pi_{ij}}\left(w_{i}\widehat{E}_{i}-w_{j}\widehat{E}_{j}\right)^{2}\]
As pointed out in \citet{Deville_Sarndal_1992}, in the definition of the variance estimators $\widehat{V}_{N,\bullet,cal,HT}^{2}$ and $\widehat{V}_{N,\bullet,cal,SYG}^{2}$ we may use the inverse inclusion probabilities $1/\pi_{i}$ in place of the sample weights $w_{i}$.

\smallskip

Finally, we need an estimator for the $g(t)$ functions defined in formulae (\ref{g_3_definition}) and (\ref{g_2_definition}) in appendix A. These functions depend on $\nu_{r}$, $\delta$, $f(\nu_{r})$, $F(\nu)$, $b_{3}$ and $b_{2}(r)$ which must be estimated from the sample. In the simulation study of this article we estimated the $g(t)$ functions by substituting plug-in estimates for $\nu_{r}$, $\delta$, $F(\nu)$, $b_{3}$ and $b_{2}(r)$. In order to estimate $f(\nu_{r})$ we used a method which is already known from \citet{Francisco_Fuller_1986}, \citet{Kovar_Rao_Wu_1988} and \citet{RKM_1990}. The method consists in equating the length of the 95\% normal confidence interval for $\nu_{r}$ to the length of the corresponding Woodruff confidence interval. The former is given by
\[2z_{0.025}\frac{\widehat{\sigma}_{\diamond}}{\widehat{f(\nu_{r})}}\]
where $\widehat{\sigma}_{\diamond}$ is an estimate for the standard deviation of the cdf-estimator $\widehat{F}_{\diamond}(t)$ at the point $t=\widehat{\nu}_{r}$ and $\widehat{f(\nu_{r})}$ is the estimate for $f(\nu_{r})$ we are looking for. On the other side of the equation, the length of the Woodruff confidence interval is given by
\[\widehat{F}^{-1}(r+z_{0.025}\widehat{\sigma}_{\diamond})- \widehat{F}^{-1}(r-z_{0.025}\widehat{\sigma}_{\diamond}).\]
Hence, the estimator for $f(\nu_{r})$ used in the simulations is given by
\begin{equation}\label{density_estimator}
\widehat{f(\nu_{r})}=\frac{2z_{0.025}\widehat{\sigma}_{\diamond}}{\widehat{F}^{-1}(r+z_{0.025}\widehat{\sigma}_{\diamond})- \widehat{F}^{-1}(r-z_{0.025}\widehat{\sigma}_{\diamond})}.
\end{equation}
Note that $\widehat{f(\nu_{r})}$ is a density estimator and hence we expect some bias from it. A formal analysis of this bias lies outside of the scope of this article. However, our simulation results suggest that is goes to zero slowly since our estimates of the relative bias of the variance estimators $\widehat{V}_{N,\bullet,Ha,SYG}^{2}$ and $\widehat{V}_{N,\bullet,cal,SYG}^{2}$ (which depend $\widehat{f(\nu_{r})}$) do not always decrease in absolute value as the sample size passes from $n=40$ to $n=80$.

As for $\widehat{\sigma}_{\diamond}$ which appears in the definition of $\widehat{f(\nu_{r})}$, in the simulations we used the square root of the Sen-Yates-Grundy estimators
\[\widehat{\sigma}_{Ha}^{2}:=-\frac{1}{2}\sum_{i\in s}\sum_{j\in s}\frac{\pi_{ij}-\pi_{i}\pi_{j}}{\pi_{ij}}\left(\frac{I(y_{i}\leq \widehat{\nu}_{r})-\widehat{F}_{Ha}(\widehat{\nu}_{r})}{\pi_{i}} - \frac{I(y_{j}\leq \widehat{\nu}_{r})-\widehat{F}_{Ha}(\widehat{\nu}_{r})}{\pi_{j}}\right)^{2}\]
and
\[\widehat{\sigma}_{cal}^{2}:=-\frac{1}{2}\sum_{i\in s}\sum_{j\in s}\frac{\pi_{ij}-\pi_{i}\pi_{j}}{\pi_{ij}}\left(w_{i}I(y_{i}\leq \widehat{\nu}_{r})- w_{j}I(y_{j}\leq \widehat{\nu}_{r})\right)^{2}.\]

\section{Simulation study}

\begin{figure}
\centering
\includegraphics[scale= 0.3]{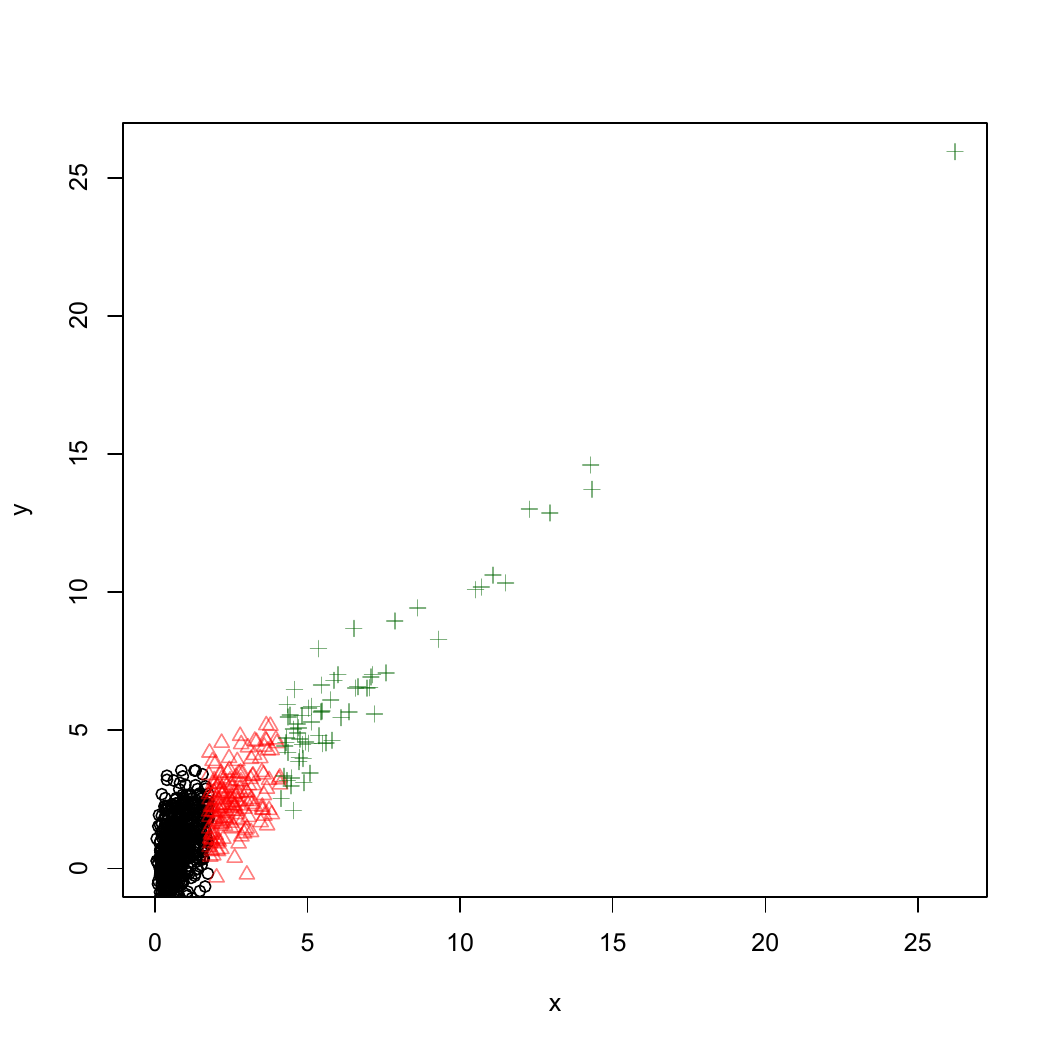}
\includegraphics[scale= 0.3]{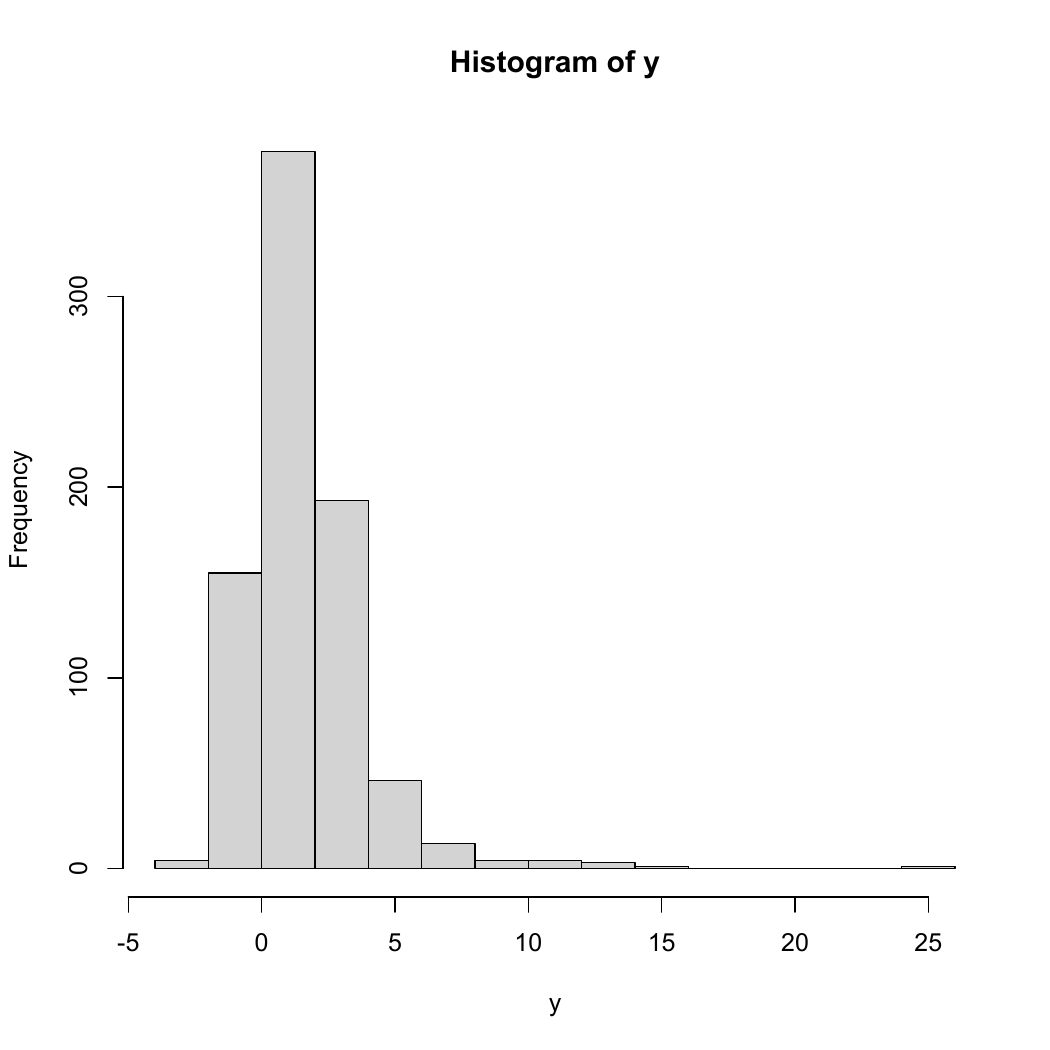}

\includegraphics[scale= 0.3]{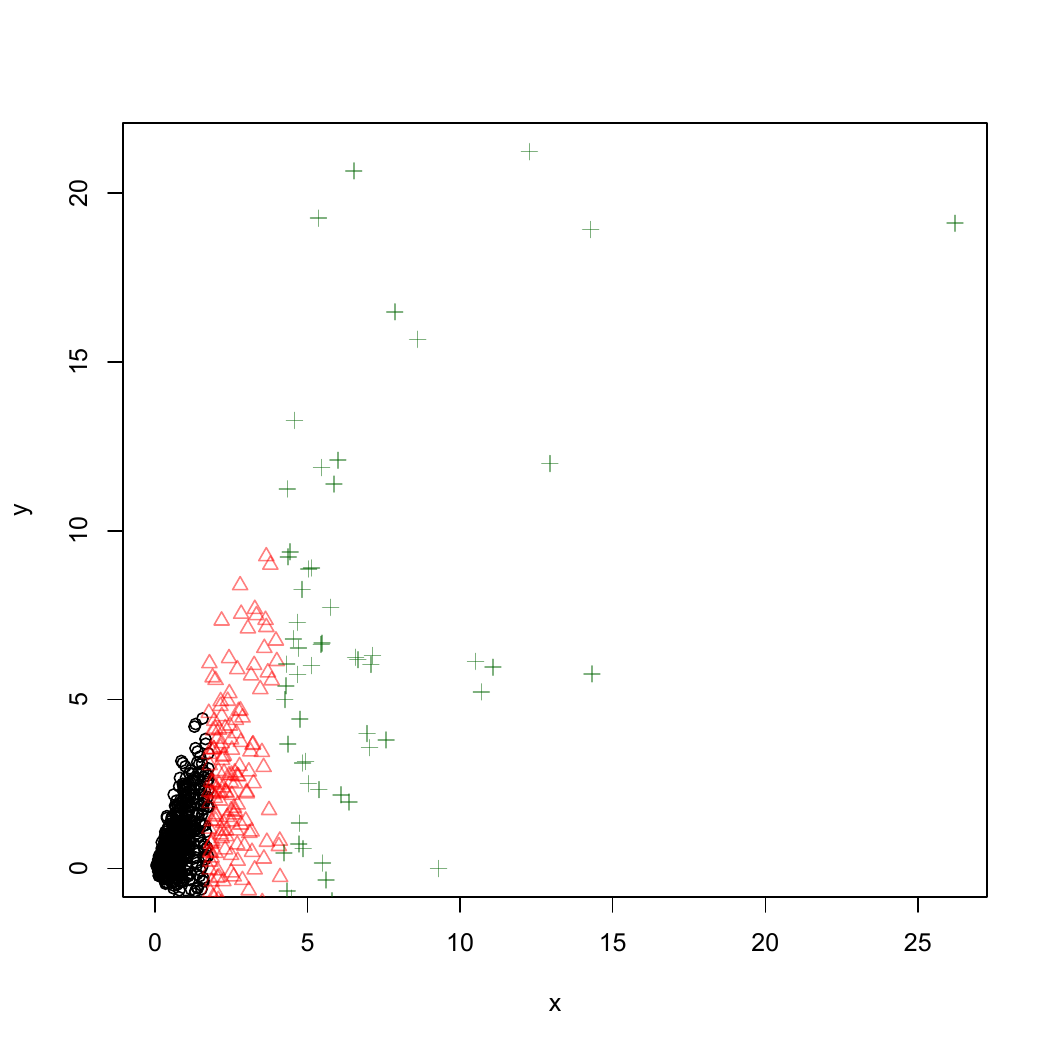}
\includegraphics[scale= 0.3]{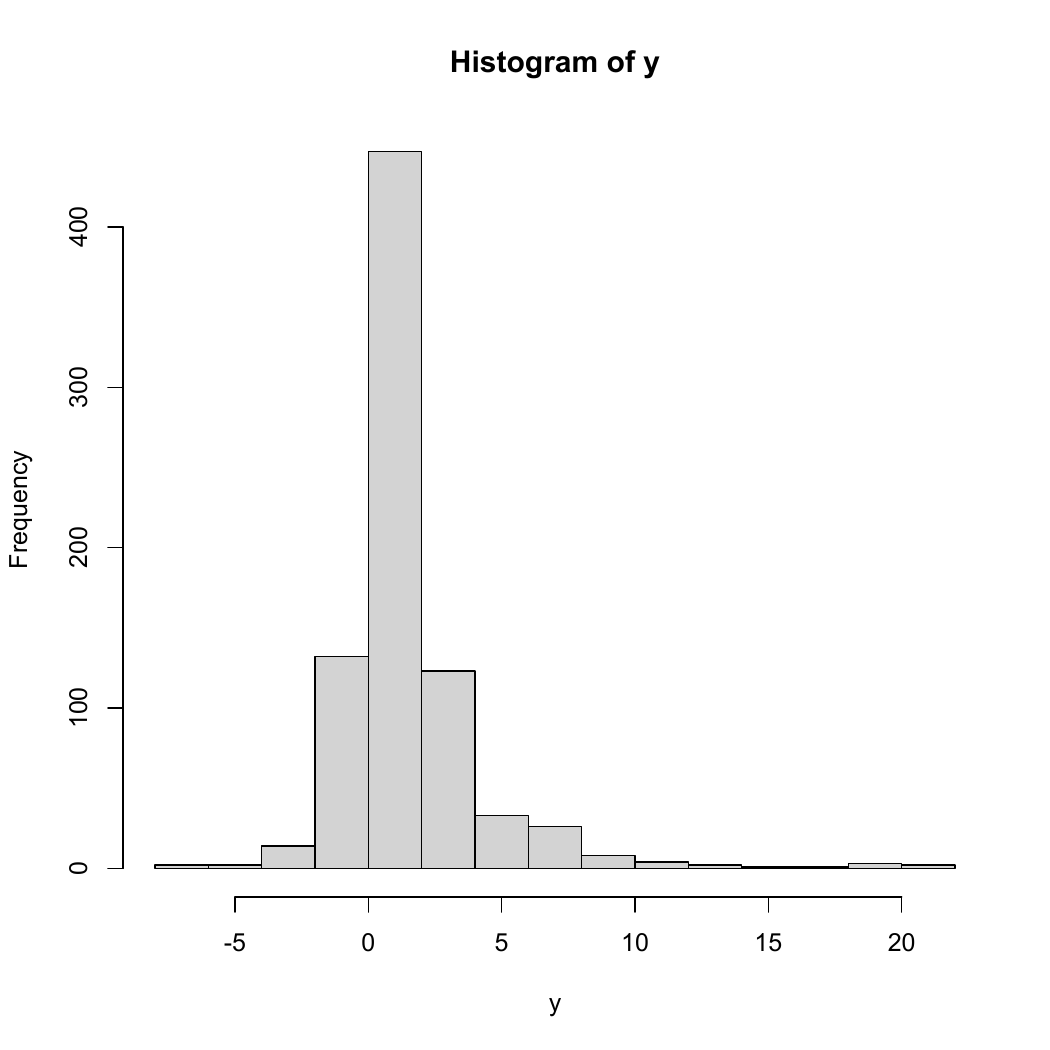}

\captionof{figure}{Population from the model with $\gamma=0$ (upper part) and $\gamma=1$ (lower part). Different symbols (circle, triangle, plus sign) are used to show  strata membership in the stratified sampling designs.}
\label{figure_populations}

\end{figure}

In the simulation study we considered two populations of size $N=800$ where the $X$ values are realizations of i.i.d. random variables with lognormal distribution with $E(\ln X)=0$ and $var(\ln X)=1$. Once we generated the $X$ values, we obtained the $Y$ values according to
\[y_{i}=x_{i}+v(x_{i})\varepsilon_{i},\quad i=1,2,\dots, N,\]
where the $\varepsilon_{i}$'s are realizations of independent standard normal random variables. For $v(x_{i})$ we used $v(x_{i}):=x_{i}^{\gamma}$ with $\gamma$ equal to either $0$ or $1$. The two populations obtained in this way are shown in Figure \ref{figure_populations}. For the population obtained from the model with $\gamma=0$ we have $b_{2}(0.25)=0.040$ and $b_{3}=0.226$, while for the population from the model with $\gamma=1$ we have $b_{2}(0.25)=0.321$ and $b_{3}=0.455$.

As for the sample design, we considered two fixed size designs each with sample size $n=40$ and $n=80$ as to achieve sampling fractions of $5\%$ and $10\%$, respectively. The two sample designs are:
\begin{itemize}
\item[i) ] simple random sampling without replacement
\item[ii) ] stratified simple random sampling with proportional allocation where the population is divided into three strata defined by intervals of $X$ such that the aggregate value of $X$ is the approximately the same in each stratum. 
\end{itemize}
For each considered population and sampling design we simulated a first set of $1000$ independent samples in order to estimate the mse of the of the estimators $\widehat{\beta}_{2, Ha}(0.75)$, $\widehat{\beta}_{2, cal}(0.75)$, $\widehat{\beta}_{3, Ha}$ and $\widehat{\beta}_{3, cal}$, i.e. in order to obtain
\[\text{mse}_{\bullet,\diamond}=\frac{1}{1000}\sum_{h=1}^{1000}\left(\widehat{b}_{\bullet}^{(h)}-b_{\bullet,\diamond}\right)^{2}\]
where $\widehat{b}_{\bullet, \diamond}^{(h)}$ is the value taken on by $\widehat{b}_{\bullet,\diamond}$ on the $h$th simulated sample. The square root of this mse is reported in the "rmse" column of the tables in appendix B. Moreover, it is also used as denominator for the computation of the bias contribution to the mse (the "bias$^{2}$/mse" column) and for the computation of the relative bias and the relative stability of the variance estimators to be defined shortly (the "rel.bias" and "rel.stab" columns). Note that for comparison the tables report also simulation results for the H\'{a}jek and calibration estimators for the mean, i.e. for the estimators
\[\overline{y}_{Ha}:=\frac{1}{\widehat{N}}\sum_{i\in s}\frac{1}{\pi_{i}}y_{i}\]
and
\[\overline{y}_{cal}:=\frac{1}{N}\sum_{i\in s}w_{i}y_{i}.\]
Apart from the mse, all other quantities needed for the construction of the tables in appendix B are computed from an independent second set of $1000$ simulated samples. In particular, given the realizations 
\[\widehat{b}_{\bullet,\diamond}^{(h)}, \widehat{V}_{\bullet,\diamond}^{(h)},\quad h=1,2,\dots, 1000,\]
for this second set of samples, we computed, for $1-\alpha=0.90, 0.95, 0.99$, the coverage rates
\[\text{coverage rate}:=\frac{1}{1000}\sum_{h=1}^{1000}I(\widehat{b}_{\bullet, \diamond}-z_{\alpha/2}\widehat{V}_{\bullet,\diamond}<b_{\bullet}<\widehat{b}_{\bullet, \diamond}+z_{\alpha/2}\widehat{V}_{\bullet,\diamond}),\]
the corresponding left and right tail errors
\[\text{left tail error LTE}:=\frac{1}{1000}\sum_{h=1}^{1000}I(b_{\bullet}<\widehat{b}_{\bullet, \diamond}-z_{\alpha/2}\widehat{V}_{\bullet,\diamond})\]
\[\text{right tail error RTE}:=\frac{1}{1000}\sum_{h=1}^{1000}I(\widehat{b}_{\bullet, \diamond}+z_{\alpha/2}\widehat{V}_{\bullet,\diamond}<b_{\bullet})\]
and 
\[\text{bias}:=\frac{1}{1000}\sum_{h=1}^{1000}\widehat{b}_{\bullet, \diamond}^{(h)}- b_{\bullet}.\]
Moreover, following \citet{Kovar_Rao_Wu_1988}, we computed
\[\text{rel.bias}(\widehat{V}_{\bullet,\diamond}^{2}):=\frac{\sum_{h=1}^{1000}\widehat{V}_{\bullet,\diamond}^{(h)2}/1000}{\text{mse}}-1\]
and
\[\text{rel.stab.}(\widehat{V}_{\bullet,\diamond}^{2}):=\frac{\sqrt{\sum_{h=1}^{1000}(\widehat{V}_{\bullet,\diamond}^{(h)2}-\text{mse}_{\bullet,\diamond})^{2}/1000}}{\text{mse}_{\bullet,\diamond}}-1,\]
where $\text{mse}_{\bullet,\diamond}$ is computed from the first set of $1000$ samples.

From the tables in appendix B we see that the coverage rates of the confidence intervals for $b_{2}(0.75)$ and $b_{3}$ are usually larger than their nominal level. In fact, in most cases the LTE and the RTE are both below their nominal level. In the samples from the populations with $\gamma=0$ we usually observe $LTE<RTE$, while in the samples from the populations with $\gamma=1$ we usually observe the opposite inequality. Also, we note little difference between the coverage accuracy of the confidence intervals based on $\widehat{b}_{\bullet,Ha}$ and those based on $\widehat{b}_{\bullet,cal}$. For comparison, the confidence intervals for the mean usually suffer from undercoverage. For the confidence intervals based on $\overline{y}_{Ha}$ we usually observe $LTE<RTE$, while the LTE and RTE of the confidence intervals based on $\overline{y}_{cal}$ are more balanced. 

Now, consider the lower parts of the tables. There we see that the rmse's of the calibration estimators $\widehat{b}_{\bullet,cal}$ are usually close to those of the H\'{a}jek estimators $\widehat{b}_{\bullet,Ha}$ suggesting that $\widehat{F}_{cal}(t)$ takes little advantage of the auxiliary information. By contrast, the rmse of $\overline{y}_{cal}$ is (as expected) constantly smaller than the rmse of $\overline{y}_{Ha}$ with large observed differences for the populations generated from the model with $\gamma=0$ (i.e. when the correlation between the $x_{i}$ and $y_{i}$ values is stronger). 

From the avg$(\widehat{V})$ column we can infer the average length of the confidence intervals. Consistently with our observations about the rmse's, we note little difference between the avg$(\widehat{V})$ values corresponding to $\widehat{b}_{\bullet,Ha}$ and $\widehat{b}_{\bullet,cal}$, while the difference between the avg$(\widehat{V})$ values corresponding to $\overline{y}_{Ha}$ and $\overline{y}_{cal}$ is often quite large especially in the cases where the population was generated from the model with $\gamma=0$. 

Finally, we analyze the relative bias and the relative stability of the variance estimators. Consistently with our conjecture about the order of the bias of $\widehat{f(\nu_{r})}$, we note that the bias of the variance estimators $\widehat{V}(\widehat{b}_{\bullet,\diamond})$ is often a considerable fraction of the mean square error of $\widehat{b}_{\bullet,\diamond}$ (for $n=80$ it is often larger than $0.30$ and in some cases the relative bias increases as the sample size passes from $n=40$ to $n=80$). To provide a benchmark value for the relative stability of the variance estimators, we note that in case of i.i.d. sampling from a normal distribution the true relative stability of the unbiased estimator for the variance of the sample mean is $\sqrt{2/(n-1)}$ which is $0.226$ for $n=40$ and $0.159$ for $n=80$. For the estimators considered in our simulation study we usually obtained relative stabilities between 2 and 4 times as large. Such multiples are in line with those that can be deduced from the simulation results in \citet*{Kovar_Rao_Wu_1988} for the several variance estimators of the median estimator considered by them (the multipliers can be obtained from Table 4 on page 40 in KRW using $n=32\times 2=64$ or $n=32\times5=160$ since KRW consider stratified sampling with $L=32$ strata and strata sample sizes of either $2$ or $5$).

\section{Conclusions and final remarks}

In this article we explored finite population inference for Bowley's skewness measure $b_{2}$ and for the Groeneveld-Meeden index $b_{3}$. In particular we considered plug-in estimators based on the H\'{a}jek cdf-estimator and on a Deville-Särndal type calibration estimator. We employed a heuristic argument to derive asymptotic variance formulae and proposed corresponding estimators. 

We tested the estimators and corresponding normal confidence intervals in a simulation study. In the simulations we usually obtained larger than nominal coverage rates for the confidence intervals. Moreover we observed little difference in terms of coverage rate, bias and rmse between the plug-in estimators based on the H\'{a}jek cdf-estimator and the plug-in estimators based on the calibration cdf-estimator. This result is likely due to the fact that the calibration cdf-estimator does not exploit the auxiliary information in efficient way. 

As for the variance estimators, we observed large bias relative to the empirical mse's of the estimators for $b_{2}$ and $b_{3}$ which is likely due to the density estimator on which the variance estimators depend. However, for srs and stratified srs the relative stability (i.e. the ratio between the empirical rmse of the variance estimator as estimator for the mse of the corresponding estimator for $b_{2}$ or $b_{3}$ and the latter in the denominator) behaves quite reasonably in line with the relative stability of the variance estimators considered by \citet*{Kovar_Rao_Wu_1988} (KRW). 

We conclude this paper with some suggestions. First, we suggest to explore the use of a variance stabilizing transformation as in \citet{Staudte_2014} which may lead to more efficient confidence intervals and/or to an improved variance estimator. Second, we note that the idea underlying the Rao-Wu modified bootstrap variance estimator \citep[see][page 30 and references therein]{Kovar_Rao_Wu_1988} can be easily adapted in order to obtain an alternative variance estimator for the plug-in estimators based on the H\'{a}jek cdf-estimator. Finally, we observe that there are several alternative cdf-estimators which, under linear superpopulation models, use auxiliary information in more efficient way than the calibration estimator considered in this article \citep*[see e.g.][]{Chambers_Dunstan_1986, RKM_1990, Rueda_2007}. The use of such estimators for estimation of $b_{2}$ and/or $b_{3}$ may give rise to more efficient plug-in estimators. However, unless one is not willing to consider only plug-in estimators based on the same cdf estimator, consistency among sample estimates of different parameters of interest may be lost.

\section*{Appendix A}

Asymptotic normality of $\widehat{b}_{3}$. For $\lambda\in[0,1]$ define $F_{\lambda}(t):=F(t)+\lambda[\widehat{F}(t)-F(t)]$, $\mu_{\lambda}:=\int t dF_{\lambda}(t)$, $\nu_{\lambda}:=F_{\lambda}^{-1}(0.5)$, 
\begin{equation*}
\begin{split}
\delta_{\lambda}&:=\int|t-\nu_{\lambda}|dF_{\lambda}(t)\\
&=\int_{-\infty}^{\nu_{\lambda}}(\nu_{\lambda}-t)dF_{\lambda}(t)+\int_{\nu_{\lambda}}^{\infty}(t-\nu_{\lambda})dF_{\lambda}(t)\\
&=\int_{-\infty}^{\nu_{\lambda}}F_{\lambda}(t)dt+\int_{\nu_{\lambda}}^{\infty}[1-F_{\lambda}(t)]dt\\
\end{split}
\end{equation*}
and $b_{3;\lambda}:=\frac{\mu_{\lambda}-\nu_{\lambda}}{\delta_{\lambda}}$. Then,
\[\frac{\partial b_{3;\lambda}}{\partial\lambda}=\frac{1}{\delta_{\lambda}}\left(\frac{\partial\mu_{\lambda}}{\partial\lambda}-\frac{\partial\nu_{\lambda}}{\partial\lambda}\right)-\frac{\mu_{\lambda}-\nu_{\lambda}}{\delta_{\lambda}^{2}}\frac{\partial\delta_{\lambda}}{\partial\lambda}\]
provided that the derivatives on the rhs exist. $\frac{\partial b_{3;\lambda}}{\partial\lambda}$ is the von Mises derivative of $b_{3}$ in direction $\widehat{F}(t)-F(t)$  \citep[see][]{vonMises_1947}. Of course 
\[\frac{\partial \mu_{\lambda}}{\partial\lambda}=\int t d\widehat{F}(t)-\int t dF(t):=\widehat{\mu}-\mu\]
always exists. In order to make sure that However, $\frac{\partial \nu_{\lambda}}{\partial\lambda}$ and $\frac{\partial \delta_{\lambda}}{\partial\lambda}$ do not necessarily exist unless we impose some condition. One such condition is
\begin{itemize}
\item[A) ] $F(t)$ and $\widehat{F}(t)$ have continuous and positive density functions.
\end{itemize}
If condition A holds we obtain
\[\frac{\partial \nu_{\lambda}}{\partial\lambda}=-\frac{\widehat{F}(\nu_{\lambda})-F(\nu_{\lambda})}{f_{\lambda}(\nu_{\lambda})},\]
where $f_{\lambda}(\nu_{\lambda})$ is the density function of $F_{\lambda}(t)$ (which must be continuous and positive by condition A) and
\begin{equation*}
\begin{split}
\frac{\partial \delta_{\lambda}}{\partial\lambda}&=\left[2F_{\lambda}(\nu_{\lambda})-1\right]\frac{\partial \nu_{\lambda}}{\partial\lambda}+\int_{-\infty}^{\nu_{\lambda}}(\widehat{F}(t)-F(t))dt-\int_{\nu_{\lambda}}^{\infty}(\widehat{F}(t)-F(t))dt\\
&=\left[2F_{\lambda}(\nu_{\lambda})-1\right]\frac{\partial \nu_{\lambda}}{\partial\lambda}+2\nu_{\lambda}[\widehat{F}(\nu_{\lambda})-F(\nu_{\lambda})]+\\
&\quad-\int_{-\infty}^{\nu_{\lambda}}td[\widehat{F}(t)-F(t)]+\int_{\nu_{\lambda}}^{\infty}td[\widehat{F}(t)-F(t)].
\end{split}
\end{equation*}
Using the notation $\widehat{b}_{3}:=b_{3;\lambda=1}$ and $b_{3}:=b_{3;\lambda=0}$, we then obtain
\begin{equation}\label{b_3_expansion}
\begin{split}
\widehat{b}_{3}&=b_{3}+\left.\frac{\partial b_{3;\lambda}}{\partial\lambda}\right|_{\lambda=0}+\int_{0}^{1}\frac{\partial^{2} b_{3;\lambda}}{\partial\lambda^{2}}(1-\lambda)d\lambda\\
&=b_{3}+\int g(t) d[\widehat{F}(t)-F(t)]+R,
\end{split}
\end{equation}
where
\begin{equation}\label{g_3_definition}
g(t):=\frac{1}{\delta}\left\{t(1-b_{3})+I(t\leq \nu)\left(\frac{1}{f(\nu)}-2\nu b_{3}+2t b_{3}\right)\right\},\quad t\in\mathbb{R},
\end{equation}
$\frac{\partial^{2} b_{3;\lambda}}{\partial\lambda^{2}}$ is a quadratic function of $\widehat{F}(\nu_{\lambda})-F(\nu_{\lambda})$ and
\[R:=\int_{0}^{1}\frac{\partial^{2} b_{3;\lambda}}{\partial\lambda^{2}}(1-\lambda)d\lambda.\]

From now on we assume that $\widehat{F}(t)$ is either $\widehat{F}_{Ha}(t)$ or $\widehat{F}_{cal}(t)$. Of course, assumption A does not hold in this case, but nevertheless we assume that the following condition, which is very similar to (\ref{b_3_expansion}), holds:  
\begin{itemize}
\item[B1) ] $N\rightarrow\infty$, $F(t)\rightarrow\int_{-\infty}^{t}f(y)dy$ and
\begin{equation*}
\begin{split}
\widehat{b}_{3}&=b_{3}+\int g(t)d[\widehat{F}(t)-F(t)]+R_{N}
\end{split}
\end{equation*}
where $R_{N}$ is asymptotically negligible with respect to the integral.
\end{itemize}
In addition to B1 we also assume that
\begin{itemize}
\item[B2) ] for $V_{N}^{2}:=var(\int g(t)d\widehat{F}(t))$,
\[\frac{1}{V_{N}}\int g(t)d[\widehat{F}(t)-F(t)]\overset{\mathcal{L}}{\longrightarrow}N(0, 1).\]
\end{itemize}
It is immediately seen that B1 and B2 imply
\[\frac{\widehat{b}_{3}-b_{3}}{V_{N}}\overset{\mathcal{L}}{\longrightarrow}N(0, 1)\]
which justifies the use of normal confidence intervals. It is also worth noting that
\begin{equation}\label{asymptotic_variance_sequence}
V_{N}^{2}:=var\left(\int g(t)d\widehat{F}(t)\right)=\frac{1}{N^{2}}var\left(\sum_{i\in s}d_{i}g(y_{i})\right)
\end{equation}
where
\begin{equation}\label{asymptotic_variance_sequence_1}
d_{i}=\begin{cases}
N/(\widehat{N}\pi_{i}) & \text{ if }\widehat{F}(t)=\widehat{F}_{Ha}(t),\\
w_{i} \text{ defined in (\ref{further_weights}) and (\ref{condizione_beta})}& \text{ if }\widehat{F}(t)=\widehat{F}_{cal}(t).\\
\end{cases}
\end{equation}

In this work we do not investigate sufficient conditions under which B1 and B2 hold. However, we conjecture that in a broad range of situations of practical interest it should be possible to prove B1 and B2 perhaps by using results from \citet{Conti_Marella_2015}, \citet{Han_Wellner_2021} and/or \citet{Dey_Chaudhuri_2024}. The simulation results in appendix B support this claim.

Of course, the above reasoning can be applied to $b_{2}(r)$ as well. For $b_{2}(r)$ we get
\begin{equation}\label{g_2_definition}
g_{2}(t):=\frac{1}{\nu_{1-r}-\nu_{r}}\left[I(t\leq\nu_{1-r})\frac{b_{2}-1}{f(\nu_{1-r})}-I(t\leq\nu_{r})\frac{1+b_{2}}{f(\nu_{r})}+2I(t\leq\nu)\frac{1}{f(\nu)}\right],
\end{equation}
where $\nu_{r}:=F^{-1}(r)$.

\citet{Groeneveld_1991} provides expressions for the influence functions of $b_{2}(r)$ and $b_{3}$. Obviously, they are closely related to our $g_{2}(t)$ and $g_{3}(t)$ functions. In fact, the relation between the influence function $IF(t;F, b_{\bullet})$ and $g_{\bullet}(t)$ is given by
\[IF(t;F, b_{\bullet}):=\int g_{\bullet}(x)d [I(x\leq t) - F(x)].\]

\section*{Appendix B}

\begin{table}[h!]
\centering
\resizebox{\textwidth}{!}{
\begin{tabular}{lcccccccccc}
\midrule 
\midrule 
 & \multicolumn{3}{c}{nominal coverage} & \multicolumn{3}{c}{left tail error} & \multicolumn{3}{c}{right tail error} \\
confidence interval & 0.90 & 0.95 & 0.99 & 0.05 & 0.025 & 0.01 & 0.05 & 0.025 & 0.01 \\
\midrule 
ci for $\mu$ based on $\widehat{F}_{Ha}$ & 0.863 & 0.914 & 0.971 & 0.032 & 0.011 & 0.003 & 0.105 & 0.075 & 0.026 \\
ci for $\mu$ based on $\widehat{F}_{cal}$ & 0.878 & 0.939 & 0.987 & 0.06 & 0.028 & 0.005 & 0.062 & 0.033 & 0.008 \\
ci for $b_{2}(0.75)$ based on $\widehat{F}_{Ha}$ & 0.955 & 0.981 & 0.995 & 0.02 & 0.007 & 0.001 & 0.025 & 0.012 & 0.004 \\
ci for $b_{2}(0.75)$ based on $\widehat{F}_{cal}$ & 0.931 & 0.967 & 0.99 & 0.025 & 0.011 & 0.003 & 0.044 & 0.022 & 0.007 \\
ci for $b_{3}$ based on $\widehat{F}_{Ha}$ & 0.915 & 0.957 & 0.986 & 0.042 & 0.021 & 0.006 & 0.043 & 0.022 & 0.008 \\
ci for $b_{3}$ based on $\widehat{F}_{cal}$ & 0.882 & 0.925 & 0.974 & 0.019 & 0.01 & 0.002 & 0.099 & 0.065 & 0.024 \\
\end{tabular}
}
\resizebox{\textwidth}{!}{
\begin{tabular}{lcccccccccc}
\midrule 
estimator & bias & rmse & bias$^{2}$/mse & avg$(\widehat{V})$ & rel.bias$(\widehat{V}^{2})$ & rel.stab.$(\widehat{V}^{2})$ \\
\midrule 
$\overline{y}_{Ha}$ & 0.007 & 0.337 & 0 & 0.321 & -0.002 & 0.759 \\
$\overline{y}_{cal}$ & 0.004 & 0.156 & 0.001 & 0.157 & 0.026 & 0.248 \\
$\widehat{b}_{2,Ha}(0.75)$ & -0.013 & 0.199 & 0.004 & 0.269 & 0.89 & 1.233 \\
$\widehat{b}_{2,cal}(0.75)$ & -0.016 & 0.202 & 0.006 & 0.242 & 0.258 & 0.458 \\
$\widehat{b}_{3,Ha}$ & 0.011 & 0.16 & 0.004 & 0.178 & 0.474 & 0.767 \\
$\widehat{b}_{3,cal}$ & -0.04 & 0.159 & 0.063 & 0.159 & 0.044 & 0.421 \\
\midrule 
\midrule 
\end{tabular}
}
\captionof{table}{Simple random sampling: $N=800$, $\gamma=0$, $n=40$}
\label{srs_N800_delta0_n40}
\end{table}
\begin{table}
\centering
\resizebox{\textwidth}{!}{
\begin{tabular}{lcccccccccc}
\midrule 
\midrule 
 & \multicolumn{3}{c}{nominal coverage} & \multicolumn{3}{c}{left tail error} & \multicolumn{3}{c}{right tail error} \\
confidence interval & 0.90 & 0.95 & 0.99 & 0.05 & 0.025 & 0.01 & 0.05 & 0.025 & 0.01 \\
\midrule 
ci for $\mu$ based on $\widehat{F}_{Ha}$ & 0.888 & 0.932 & 0.98 & 0.028 & 0.012 & 0.003 & 0.084 & 0.056 & 0.017 \\
ci for $\mu$ based on $\widehat{F}_{cal}$ & 0.898 & 0.942 & 0.98 & 0.06 & 0.038 & 0.009 & 0.042 & 0.02 & 0.011 \\
ci for $b_{2}(0.75)$ based on $\widehat{F}_{Ha}$ & 0.947 & 0.976 & 0.995 & 0.024 & 0.012 & 0.002 & 0.029 & 0.012 & 0.003 \\
ci for $b_{2}(0.75)$ based on $\widehat{F}_{cal}$ & 0.925 & 0.966 & 0.993 & 0.033 & 0.015 & 0.004 & 0.042 & 0.019 & 0.003 \\
ci for $b_{3}$ based on $\widehat{F}_{Ha}$ & 0.901 & 0.961 & 0.994 & 0.047 & 0.02 & 0.003 & 0.052 & 0.019 & 0.003 \\
ci for $b_{3}$ based on $\widehat{F}_{cal}$ & 0.905 & 0.945 & 0.987 & 0.025 & 0.012 & 0.003 & 0.07 & 0.043 & 0.01 \\
\end{tabular}
}
\resizebox{\textwidth}{!}{
\begin{tabular}{lcccccccccc}
\midrule 
estimator & bias & rmse & bias$^{2}$/mse & avg$(\widehat{V})$ & rel.bias$(\widehat{V}^{2})$ & rel.stab.$(\widehat{V}^{2})$ \\
\midrule 
$\overline{y}_{Ha}$ & 0.012 & 0.24 & 0.002 & 0.228 & -0.042 & 0.505 \\
$\overline{y}_{cal}$ & 0.007 & 0.113 & 0.004 & 0.109 & -0.065 & 0.16 \\
$\widehat{b}_{2,Ha}(0.75)$ & -0.01 & 0.143 & 0.005 & 0.165 & 0.355 & 0.55 \\
$\widehat{b}_{2,cal}(0.75)$ & -0.011 & 0.143 & 0.006 & 0.156 & 0.209 & 0.427 \\
$\widehat{b}_{3,Ha}$ & 0.001 & 0.109 & 0 & 0.119 & 0.232 & 0.49 \\
$\widehat{b}_{3,cal}$ & -0.021 & 0.1 & 0.046 & 0.107 & 0.206 & 0.559 \\
\midrule 
\midrule 
\end{tabular}
}
\captionof{table}{Simple random sampling: $N=800$, $\gamma=0$, $n=80$}
\label{srs_N800_delta0_n80}
\end{table}

\begin{table}
\centering
\resizebox{\textwidth}{!}{
\begin{tabular}{lcccccccccc}
\midrule 
\midrule 
 & \multicolumn{3}{c}{nominal coverage} & \multicolumn{3}{c}{left tail error} & \multicolumn{3}{c}{right tail error} \\
confidence interval & 0.90 & 0.95 & 0.99 & 0.05 & 0.025 & 0.01 & 0.05 & 0.025 & 0.01 \\
\midrule 
ci for $\mu$ based on $\widehat{F}_{Ha}$ & 0.846 & 0.903 & 0.957 & 0.017 & 0.005 & 0 & 0.137 & 0.092 & 0.043 \\
ci for $\mu$ based on $\widehat{F}_{cal}$ & 0.821 & 0.886 & 0.949 & 0.081 & 0.049 & 0.024 & 0.098 & 0.065 & 0.027 \\
ci for $b_{2}(0.75)$ based on $\widehat{F}_{Ha}$ & 0.969 & 0.987 & 0.998 & 0.021 & 0.009 & 0.002 & 0.01 & 0.004 & 0 \\
ci for $b_{2}(0.75)$ based on $\widehat{F}_{cal}$ & 0.946 & 0.971 & 0.992 & 0.039 & 0.025 & 0.007 & 0.015 & 0.004 & 0.001 \\
ci for $b_{3}$ based on $\widehat{F}_{Ha}$ & 0.916 & 0.953 & 0.993 & 0.055 & 0.035 & 0.006 & 0.029 & 0.012 & 0.001 \\
ci for $b_{3}$ based on $\widehat{F}_{cal}$ & 0.882 & 0.934 & 0.984 & 0.054 & 0.03 & 0.008 & 0.064 & 0.036 & 0.008 \\
\end{tabular}
}
\resizebox{\textwidth}{!}{
\begin{tabular}{lcccccccccc}
\midrule 
estimator & bias & rmse & bias$^{2}$/mse & avg$(\widehat{V})$ & rel.bias$(\widehat{V}^{2})$ & rel.stab.$(\widehat{V}^{2})$ \\
\midrule 
$\overline{y}_{Ha}$ & -0.004 & 0.403 & 0 & 0.395 & 0.061 & 0.703 \\
$\overline{y}_{cal}$ & 0 & 0.355 & 0 & 0.316 & -0.146 & 0.55 \\
$\widehat{b}_{2,Ha}(0.75)$ & -0.026 & 0.199 & 0.017 & 0.291 & 1.363 & 2.467 \\
$\widehat{b}_{2,cal}(0.75)$ & -0.021 & 0.198 & 0.011 & 0.251 & 0.255 & 0.506 \\
$\widehat{b}_{3,Ha}$ & 0.005 & 0.153 & 0.001 & 0.168 & 0.777 & 1.748 \\
$\widehat{b}_{3,cal}$ & -0.031 & 0.162 & 0.037 & 0.16 & 0.02 & 0.58 \\
\midrule 
\midrule 
\end{tabular}
}
\captionof{table}{Simple random sampling: $N=800$, $\gamma=1$, $n=40$}
\label{srs_N800_delta1_n40}
\end{table}
\begin{table}
\centering
\resizebox{\textwidth}{!}{
\begin{tabular}{lcccccccccc}
\midrule 
\midrule 
 & \multicolumn{3}{c}{nominal coverage} & \multicolumn{3}{c}{left tail error} & \multicolumn{3}{c}{right tail error} \\
confidence interval & 0.90 & 0.95 & 0.99 & 0.05 & 0.025 & 0.01 & 0.05 & 0.025 & 0.01 \\
\midrule 
ci for $\mu$ based on $\widehat{F}_{Ha}$ & 0.878 & 0.928 & 0.971 & 0.026 & 0.007 & 0 & 0.096 & 0.065 & 0.029 \\
ci for $\mu$ based on $\widehat{F}_{cal}$ & 0.862 & 0.928 & 0.972 & 0.062 & 0.034 & 0.016 & 0.076 & 0.038 & 0.012 \\
ci for $b_{2}(0.75)$ based on $\widehat{F}_{Ha}$ & 0.926 & 0.963 & 0.994 & 0.046 & 0.024 & 0.006 & 0.028 & 0.013 & 0 \\
ci for $b_{2}(0.75)$ based on $\widehat{F}_{cal}$ & 0.918 & 0.958 & 0.992 & 0.05 & 0.027 & 0.008 & 0.032 & 0.015 & 0 \\
ci for $b_{3}$ based on $\widehat{F}_{Ha}$ & 0.893 & 0.938 & 0.986 & 0.069 & 0.043 & 0.01 & 0.038 & 0.019 & 0.004 \\
ci for $b_{3}$ based on $\widehat{F}_{cal}$ & 0.9 & 0.944 & 0.981 & 0.047 & 0.03 & 0.011 & 0.053 & 0.026 & 0.008 \\
\end{tabular}
}
\resizebox{\textwidth}{!}{
\begin{tabular}{lcccccccccc}
\midrule 
estimator & bias & rmse & bias$^{2}$/mse & avg$(\widehat{V})$ & rel.bias$(\widehat{V}^{2})$ & rel.stab.$(\widehat{V}^{2})$ \\
\midrule 
$\overline{y}_{Ha}$ & 0.005 & 0.291 & 0 & 0.28 & -0.023 & 0.444 \\
$\overline{y}_{cal}$ & 0.003 & 0.259 & 0 & 0.227 & -0.195 & 0.458 \\
$\widehat{b}_{2,Ha}(0.75)$ & -0.01 & 0.144 & 0.005 & 0.165 & 0.37 & 0.773 \\
$\widehat{b}_{2,cal}(0.75)$ & -0.002 & 0.139 & 0 & 0.151 & 0.045 & 0.308 \\
$\widehat{b}_{3,Ha}$ & 0.004 & 0.11 & 0.002 & 0.111 & 0.223 & 0.543 \\
$\widehat{b}_{3,cal}$ & -0.013 & 0.113 & 0.013 & 0.109 & -0.041 & 0.312 \\
\midrule 
\midrule 
\end{tabular}
}
\captionof{table}{Simple random sampling: $N=800$, $\gamma=1$, $n=80$}
\label{srs_N800_delta1_n80}
\end{table}

\begin{table}
\centering
\resizebox{\textwidth}{!}{
\begin{tabular}{lcccccccccc}
\midrule 
\midrule 
 & \multicolumn{3}{c}{nominal coverage} & \multicolumn{3}{c}{left tail error} & \multicolumn{3}{c}{right tail error} \\
confidence interval & 0.90 & 0.95 & 0.99 & 0.05 & 0.025 & 0.01 & 0.05 & 0.025 & 0.01 \\
\midrule 
ci for $\mu$ based on $\widehat{F}_{Ha}$ & 0.873 & 0.932 & 0.984 & 0.04 & 0.026 & 0.003 & 0.087 & 0.042 & 0.013 \\
ci for $\mu$ based on $\widehat{F}_{cal}$ & 0.895 & 0.938 & 0.98 & 0.058 & 0.037 & 0.014 & 0.047 & 0.025 & 0.006 \\
ci for $b_{2}(0.75)$ based on $\widehat{F}_{Ha}$ & 0.943 & 0.971 & 0.996 & 0.021 & 0.008 & 0.001 & 0.036 & 0.021 & 0.003 \\
ci for $b_{2}(0.75)$ based on $\widehat{F}_{cal}$ & 0.931 & 0.968 & 0.996 & 0.027 & 0.012 & 0.002 & 0.042 & 0.02 & 0.002 \\
ci for $b_{3}$ based on $\widehat{F}_{Ha}$ & 0.924 & 0.965 & 0.989 & 0.021 & 0.013 & 0.002 & 0.055 & 0.022 & 0.009 \\
ci for $b_{3}$ based on $\widehat{F}_{cal}$ & 0.918 & 0.956 & 0.987 & 0.018 & 0.009 & 0.002 & 0.064 & 0.035 & 0.011 \\
\end{tabular}
}
\resizebox{\textwidth}{!}{
\begin{tabular}{lcccccccccc}
\midrule 
estimator & bias & rmse & bias$^{2}$/mse & avg$(\widehat{V})$ & rel.bias$(\widehat{V}^{2})$ & rel.stab.$(\widehat{V}^{2})$ \\
\midrule 
$\overline{y}_{Ha}$ & -0.004 & 0.226 & 0 & 0.21 & -0.056 & 0.722 \\
$\overline{y}_{cal}$ & 0.002 & 0.163 & 0 & 0.158 & -0.052 & 0.207 \\
$\widehat{b}_{2,Ha}(0.75)$ & -0.016 & 0.2 & 0.006 & 0.237 & 0.44 & 0.658 \\
$\widehat{b}_{2,cal}(0.75)$ & -0.014 & 0.201 & 0.005 & 0.239 & 0.367 & 0.625 \\
$\widehat{b}_{3,Ha}$ & -0.01 & 0.139 & 0.006 & 0.16 & 0.46 & 0.705 \\
$\widehat{b}_{3,cal}$ & -0.026 & 0.132 & 0.04 & 0.151 & 0.369 & 0.66 \\
\midrule 
\midrule 
\end{tabular}
}
\captionof{table}{Stratified simple random sampling: $N=800$, $\gamma=0$, $n=40$}
\label{stratified_N800_delta0_n40}
\end{table}
\begin{table}
\centering
\resizebox{\textwidth}{!}{
\begin{tabular}{lcccccccccc}
\midrule 
\midrule 
 & \multicolumn{3}{c}{nominal coverage} & \multicolumn{3}{c}{left tail error} & \multicolumn{3}{c}{right tail error} \\
confidence interval & 0.90 & 0.95 & 0.99 & 0.05 & 0.025 & 0.01 & 0.05 & 0.025 & 0.01 \\
\midrule 
ci for $\mu$ based on $\widehat{F}_{Ha}$ & 0.888 & 0.944 & 0.985 & 0.036 & 0.015 & 0.002 & 0.076 & 0.041 & 0.013 \\
ci for $\mu$ based on $\widehat{F}_{cal}$ & 0.902 & 0.962 & 0.994 & 0.049 & 0.021 & 0.003 & 0.049 & 0.017 & 0.003 \\
ci for $b_{2}(0.75)$ based on $\widehat{F}_{Ha}$ & 0.939 & 0.964 & 0.991 & 0.029 & 0.015 & 0.002 & 0.032 & 0.021 & 0.007 \\
ci for $b_{2}(0.75)$ based on $\widehat{F}_{cal}$ & 0.918 & 0.957 & 0.991 & 0.032 & 0.016 & 0.003 & 0.05 & 0.027 & 0.006 \\
ci for $b_{3}$ based on $\widehat{F}_{Ha}$ & 0.927 & 0.964 & 0.992 & 0.033 & 0.012 & 0.002 & 0.04 & 0.024 & 0.006 \\
ci for $b_{3}$ based on $\widehat{F}_{cal}$ & 0.918 & 0.96 & 0.989 & 0.023 & 0.007 & 0.002 & 0.059 & 0.033 & 0.009 \\
\end{tabular}
}
\resizebox{\textwidth}{!}{
\begin{tabular}{lcccccccccc}
\midrule 
estimator & bias & rmse & bias$^{2}$/mse & avg$(\widehat{V})$ & rel.bias$(\widehat{V}^{2})$ & rel.stab.$(\widehat{V}^{2})$ \\
\midrule 
$\overline{y}_{Ha}$ & 0.006 & 0.153 & 0.002 & 0.154 & 0.089 & 0.69 \\
$\overline{y}_{cal}$ & 0.001 & 0.111 & 0 & 0.109 & -0.026 & 0.146 \\
$\widehat{b}_{2,Ha}(0.75)$ & -0.013 & 0.138 & 0.009 & 0.156 & 0.321 & 0.556 \\
$\widehat{b}_{2,cal}(0.75)$ & -0.013 & 0.14 & 0.008 & 0.156 & 0.411 & 0.662 \\
$\widehat{b}_{3,Ha}$ & 0.002 & 0.094 & 0.001 & 0.11 & 0.267 & 0.492 \\
$\widehat{b}_{3,cal}$ & -0.019 & 0.09 & 0.044 & 0.101 & 0.316 & 0.617 \\
\midrule 
\midrule 
\end{tabular}
}
\captionof{table}{Stratified simple random sampling: $N=800$, $\gamma=0$, $n=80$}
\label{stratified_N800_delta0_n80}
\end{table}
\begin{table}
\centering
\resizebox{\textwidth}{!}{
\begin{tabular}{lcccccccccc}
\midrule 
\midrule 
 & \multicolumn{3}{c}{nominal coverage} & \multicolumn{3}{c}{left tail error} & \multicolumn{3}{c}{right tail error} \\
confidence interval & 0.90 & 0.95 & 0.99 & 0.05 & 0.025 & 0.01 & 0.05 & 0.025 & 0.01 \\
\midrule 
ci for $\mu$ based on $\widehat{F}_{Ha}$ & 0.883 & 0.929 & 0.979 & 0.047 & 0.026 & 0.007 & 0.07 & 0.045 & 0.014 \\
ci for $\mu$ based on $\widehat{F}_{cal}$ & 0.873 & 0.916 & 0.977 & 0.062 & 0.043 & 0.014 & 0.065 & 0.041 & 0.009 \\
ci for $b_{2}(0.75)$ based on $\widehat{F}_{Ha}$ & 0.951 & 0.977 & 0.994 & 0.031 & 0.015 & 0.006 & 0.018 & 0.008 & 0 \\
ci for $b_{2}(0.75)$ based on $\widehat{F}_{cal}$ & 0.958 & 0.986 & 0.995 & 0.029 & 0.01 & 0.005 & 0.013 & 0.004 & 0 \\
ci for $b_{3}$ based on $\widehat{F}_{Ha}$ & 0.908 & 0.962 & 0.987 & 0.049 & 0.023 & 0.008 & 0.043 & 0.015 & 0.005 \\
ci for $b_{3}$ based on $\widehat{F}_{cal}$ & 0.911 & 0.96 & 0.988 & 0.04 & 0.017 & 0.007 & 0.049 & 0.023 & 0.005 \\
\end{tabular}
}
\resizebox{\textwidth}{!}{
\begin{tabular}{lcccccccccc}
\midrule 
estimator & bias & rmse & bias$^{2}$/mse & avg$(\widehat{V})$ & rel.bias$(\widehat{V}^{2})$ & rel.stab.$(\widehat{V}^{2})$ \\
\midrule 
$\overline{y}_{Ha}$ & -0.003 & 0.334 & 0 & 0.329 & 0.023 & 0.49 \\
$\overline{y}_{cal}$ & 0.007 & 0.327 & 0 & 0.32 & 0.016 & 0.519 \\
$\widehat{b}_{2,Ha}(0.75)$ & -0.016 & 0.207 & 0.006 & 0.25 & 0.59 & 1.373 \\
$\widehat{b}_{2,cal}(0.75)$ & -0.018 & 0.205 & 0.008 & 0.251 & 0.151 & 0.462 \\
$\widehat{b}_{3,Ha}$ & -0.004 & 0.149 & 0.001 & 0.157 & 0.633 & 1.428 \\
$\widehat{b}_{3,cal}$ & -0.017 & 0.148 & 0.014 & 0.157 & 0.174 & 0.5 \\
\midrule 
\midrule 
\end{tabular}
}
\captionof{table}{Stratified simple random sampling: $N=800$, $\gamma=1$, $n=40$}
\label{stratified_N800_delta1_n40}
\end{table}
\begin{table}
\centering
\resizebox{\textwidth}{!}{
\begin{tabular}{lcccccccccc}
\midrule 
\midrule 
 & \multicolumn{3}{c}{nominal coverage} & \multicolumn{3}{c}{left tail error} & \multicolumn{3}{c}{right tail error} \\
confidence interval & 0.90 & 0.95 & 0.99 & 0.05 & 0.025 & 0.01 & 0.05 & 0.025 & 0.01 \\
\midrule 
ci for $\mu$ based on $\widehat{F}_{Ha}$ & 0.877 & 0.934 & 0.983 & 0.046 & 0.025 & 0.006 & 0.077 & 0.041 & 0.011 \\
ci for $\mu$ based on $\widehat{F}_{cal}$ & 0.878 & 0.945 & 0.984 & 0.052 & 0.027 & 0.007 & 0.07 & 0.028 & 0.009 \\
ci for $b_{2}(0.75)$ based on $\widehat{F}_{Ha}$ & 0.928 & 0.966 & 0.992 & 0.044 & 0.023 & 0.007 & 0.028 & 0.011 & 0.001 \\
ci for $b_{2}(0.75)$ based on $\widehat{F}_{cal}$ & 0.934 & 0.965 & 0.994 & 0.039 & 0.022 & 0.005 & 0.027 & 0.013 & 0.001 \\
ci for $b_{3}$ based on $\widehat{F}_{Ha}$ & 0.901 & 0.943 & 0.99 & 0.056 & 0.032 & 0.006 & 0.043 & 0.025 & 0.004 \\
ci for $b_{3}$ based on $\widehat{F}_{cal}$ & 0.911 & 0.95 & 0.992 & 0.045 & 0.028 & 0.004 & 0.044 & 0.022 & 0.004 \\
\end{tabular}
}
\resizebox{\textwidth}{!}{
\begin{tabular}{lcccccccccc}
\midrule 
estimator & bias & rmse & bias$^{2}$/mse & avg$(\widehat{V})$ & rel.bias$(\widehat{V}^{2})$ & rel.stab.$(\widehat{V}^{2})$ \\
\midrule 
$\overline{y}_{Ha}$ & -0.003 & 0.246 & 0 & 0.239 & -0.031 & 0.321 \\
$\overline{y}_{cal}$ & -0.005 & 0.239 & 0.001 & 0.227 & -0.07 & 0.355 \\
$\widehat{b}_{2,Ha}(0.75)$ & -0.015 & 0.143 & 0.011 & 0.153 & 0.196 & 0.555 \\
$\widehat{b}_{2,cal}(0.75)$ & -0.01 & 0.14 & 0.006 & 0.151 & -0.004 & 0.287 \\
$\widehat{b}_{3,Ha}$ & -0.008 & 0.11 & 0.005 & 0.108 & 0.209 & 0.568 \\
$\widehat{b}_{3,cal}$ & -0.015 & 0.107 & 0.018 & 0.108 & 0.023 & 0.296 \\
\midrule 
\midrule 
\end{tabular}
}
\captionof{table}{Stratified simple random sampling: $N=800$, $\gamma=1$, $n=80$}
\label{stratified_N800_delta1_n80}
\end{table}

\newpage

\bibliographystyle{apalike}
\bibliographystyle{plain}
\bibliographystyle{plainnat}
\bibliography{inference_for_skewness_indices_nuovo_1}

\end{document}